\documentclass{article}

\usepackage{arxiv}

\usepackage[utf8]{inputenc} 
\usepackage[T1]{fontenc}    
\usepackage{hyperref}       
\usepackage{url}            
\usepackage{booktabs}       
\usepackage{amsfonts}       
\usepackage{nicefrac}       
\usepackage{microtype}      
\usepackage{lipsum}

\usepackage{relsize}
\usepackage{verbatim}
\usepackage{graphicx}
\graphicspath{{Figures_eps/}}
\usepackage{dcolumn}
\usepackage{bm}
\usepackage{subcaption}
\usepackage{float}

\usepackage{amsmath,amssymb,amsfonts}
\usepackage{graphics}
\usepackage{color}
\usepackage{xcolor}
\definecolor{rev}{rgb}{0,0,0}

\usepackage{cite}

\usepackage[ruled,vlined]{algorithm2e}
\title{Multifidelity Computing for Coupling Full and Reduced Order Models}

\author{
 Shady E. Ahmed \\
  School of Mechanical \& Aerospace Engineering,\\
  Oklahoma State University,\\
  Stillwater, OK 74078, USA.\\
  \texttt{shady.ahmed@okstate.edu}
\And
 Omer San \\
  School of Mechanical \& Aerospace Engineering,\\
  Oklahoma State University,\\
  Stillwater, OK 74078, USA.\\
  \texttt{osan@okstate.edu} 
\And
 Kursat Kara \\
  School of Mechanical \& Aerospace Engineering,\\
  Oklahoma State University,\\
  Stillwater, OK 74078, USA.\\
  \texttt{kursat.kara@okstate.edu} 
\And
Rami Younis\\
The McDougall School of Petroleum Engineering,\\
The University of Tulsa,\\
Tulsa, OK 74104, USA.\\
\texttt{rami-younis@utulsa.edu}
\And
Adil Rasheed\\
Department of Engineering Cybernetics,\\
Norwegian University of Science and Technology,\\
N-7465, Trondheim, Norway.\\
\texttt{adil.rasheed@ntnu.no }
}

\begin{document}
\maketitle

\begin{abstract}
Hybrid physics-machine learning models are increasingly being used in simulations of transport processes. Many complex multiphysics systems relevant to scientific and engineering applications include multiple spatiotemporal scales and comprise a multifidelity problem sharing an interface between various formulations or heterogeneous computational entities. To this end, we present a robust hybrid analysis and modeling approach combining a physics-based full order model (FOM) and a data-driven reduced order model (ROM) to form the building blocks of an integrated approach among mixed fidelity descriptions toward predictive digital twin technologies. At the interface, we introduce a long short-term memory network to bridge these high and low-fidelity models in various forms of interfacial error correction or prolongation. The proposed interface learning approaches are tested as a new way to address ROM-FOM coupling problems solving nonlinear advection-diffusion flow situations with a bifidelity setup that captures the essence of a broad class of transport processes. 
\end{abstract}


\section{Introduction}
Numerical simulations are the workhorse for the design, testing, and implementation of scientific infrastructure and engineering applications. While immense advances in computational mathematics and scientific computing have come to fruition, such simulations usually suffer a curse of dimensionality limiting turnaround. The field of multifidelity computing, therefore, aims to address this computational challenge by exploiting the relationship between high-fidelity and low-fidelity models. One such multifidelity approach becomes crucial, especially for multi-query applications, such as optimization, inference, and uncertainty quantification, that require multiple model evaluations in an outer-workflow loop. To this end, sampling-based approaches have been often introduced to leverage information from many evaluations of inexpensive low-fidelity models fused by only a few carefully selected high-fidelity computations. An excellent review of the state-of-the-art multifidelity approaches for outer-loop contexts can be found in \cite{peherstorfer2018survey}. 

In this paper, we focus on a different type of multifidelity formulation targeting domain decomposition type problems that consist of multiple zones with different characteristics \textcolor{rev}{as well as multiphysics systems where different levels of solvers are devoted to coupled physical phenomena.} A key aspect of the zonal multifidelity approach is its ability to handle intrinsic heterogeneous physical properties, varying geometries, and underlying governing dynamics. This heterogeneity can be mild as in aerospace applications with spatially varying parameters. However, in media where there is a permittivity such as in electrostatics or porous media, this might be more pronounced. For example, fluid flow in rock often follows Darcy's law, whereas flow in a fracture is modeled as Poiseuille flow. Moreover, a related process in subsurface flows might include a high fidelity approach around wells (that drive the flow) and a low-fidelity model for subdomains in the interior \cite{bjorkevoll2015use,macpherson2015technology}. This discussion can also be extended to an active flow control problem to elucidate the concept of the \emph{zonal multifidelity approach} that we tackle in this work.  In general, boundary-layer control poses a grand challenge in many aerospace applications including lift enhancement, noise mitigation, transition delay, and drag reduction. Among many other actuator technologies, blooming jets \cite{reynolds2003bifurcating,tyliszczak2015multi,gohil2015simulation} and sweeping jets \cite{phillips2013use,koklu2016effect,koklu2018effects} offer new prospective solutions in improving the aerodynamics efficiency and performance of the future air vehicle systems. The size of these actuators is usually orders of magnitude smaller than the length scales of the entire computational domain (e.g., an aircraft wing or tail). Including the full representations of each controller’s internal flow dynamics in a comprehensive numerical analysis of the entire system is an extremely daunting approach \cite{kara2018flow}. Meanwhile, the effective flow physics of these actuators can often be accurately characterized by a latent reduced order space due to the existence of strong coherent structures such as quasi-periodic or time-periodic shedding, pulsation, or jet actuation. Therefore, in practice, those flow actuators can be modeled by considering a reduced order surrogate coupled and tied to the global simulation of the whole wing or tail \cite{childs2016simulation,aram2019synchronization}.

The above examples illustrate that different levels of models and descriptions can be devoted to different zones and components of the problem in order to allocate computational resources more effectively and economically. This might be the case for many other coupled multiphysics systems, such as geometric multiscale \cite{quarteroni2003analysis,d2008coupling,passerini20093d,quarteroni2016geometric,boon2018robust,nordbotten2019unified} and heterogeneous multiscale \cite{kevrekidis2009equation,weinan2003heterogeneous} problems, fluid-structure interactions \cite{van2011partitioned}, and industrial scale applications \cite{shankaran2001multi,xu2020reduced,siddiqui2019numerical}. Since various zones and/or physics in these systems are connected through interfaces, data sharing, and consistent interface treatment among respective models are inevitable. Likewise, multirate and locally adaptive stepping methods can yield a mismatch at the space-time interface, and simple interpolation or extrapolation might lead to solution divergence or instabilities \cite{gander2013techniques}. Moreover, even if we are interested in simulating just one portion of the domain corresponding to some specific dynamics, we still need to specify the physically consistent interface conditions. Running a high fidelity solver everywhere only to provide the flow state at the interface seems to be unreasonable. Therefore, we consider formulating an interface modeling approach to facilitate the development of efficient and reliable multifidelity computing. This should serve and advance the applicability of the emerging digital twin technologies in many sectors \cite{rasheed2020digital}. However, just like any technology, it comes with its own needs and challenges \cite{tao2018digital,hartmann2018model,ganguli2020digital,chakraborty2020machine,kapteyn2020data,kapteyn2020physics}. In practice, two modeling paradigms are in order.


\begin{itemize}
\item \emph{Physics-based modeling:} This approach involves careful observation of a physical phenomenon of interest, development of its partial understanding, expression of the understanding in the form of mathematical equations, and ultimately, solution of these equations. Due to the partial understanding and numerous assumptions along the steps from observation to the solution of the equations, a large portion of the essential governing physics might be, intentionally or unintentionally, ignored. The applicability of high fidelity simulators with minimal assumptions has so far been limited to the offline design phase only. Despite this significant drawback, what makes these models attractive are sound foundations from first principles, interpretability, generalizability, and existence of robust theories for the analysis of stability and uncertainty. However, most of these models are generally computationally expensive, do not adapt to new scenarios automatically, and can be susceptible to numerical instabilities.

\item \emph{Data-driven modeling:} With the abundant supply of big data, open-source cutting edge and easy-to-use machine learning libraries, cheap computational infrastructure, and high quality, readily available training resources, data-driven modeling has become very popular. Compared to the physics-based modeling approach, these models thrive on the assumption that data is a manifestation of both known and unknown physics and hence when trained with an ample amount of data, the data-driven models might learn the full physics on their own. This approach, involving in particular deep learning, has started achieving human-level performance in several tasks that were until recently considered impossible for computers. Notable among these are image classification \cite{szegedy2017inception}, dimensionality reduction \cite{Hinton504}, medical treatment \cite{LIU20191}, smart agriculture \cite{BU2019500}, physical sciences \cite{brunton2020machine,reichstein2019deep,schmidt2019recent} and beyond. Some of the advantages of these models are online learning capability, computational efficiency for inference, accuracy even for very challenging problems as far as the training, validation and test data are prepared properly. However, due to their data-hungry and black-box nature, poor generalizability, inherent bias and lack of robust theory for the analysis of model stability, their acceptability in high stake applications like digital twin and autonomous systems is fairly limited. In fact, the numerous vulnerabilities of deep neural networks have been exposed beyond doubt in several recent works \cite{akhtar2018threat,yuan2019adversarial,hao2020adversarial}. 
\end{itemize}

\begin{figure}[ht]
\centering
\includegraphics[trim= 0 0 0 0, clip, width=0.8\linewidth]{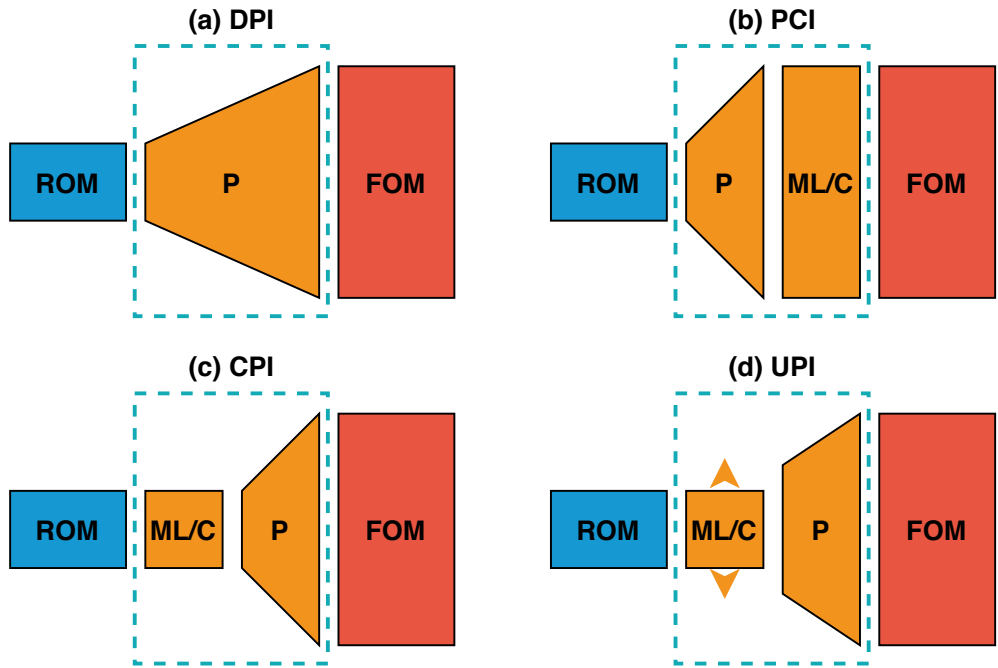}
\caption{The proposed multifidelity concepts toward hybrid ROM/FOM coupling. Dashed blocks refer to the interface learning approaches introduced in the present work: (a) Direct Prolongation Interface (DPI), (b) Prolongation followed by a machine learning Correction Interface (PCI), (c) machine learning Correction followed by a Prolongation Interface (CPI), and (d) Uplifted Prolongation Interface (UPI) where the latent space is enhanced through machine learning before we apply the prolongation operator.}
\label{fig:interface}
\end{figure}

In this work, we put forth a \emph{hybrid analysis and modeling (HAM)} framework as a new paradigm in modeling and simulations by promoting the strengths and mitigating the weaknesses of physics-driven and data-driven modeling approaches. Our HAM approach combines the generalizability, interpretability, robust foundation and understanding of physics-based modeling with the accuracy, computational efficiency, and automatic pattern-identification capabilities of advanced data-driven modeling technologies. In the context of multifidelity computing, we advocate and explore the utilization of statistical inference to bridge low-fidelity and high-fidelity descriptions. In particular, we adopt the long short-term memory (LSTM) neural network to match the reduced order model (ROM) and full order model (FOM) solutions at their intersect. To form the building blocks of our HAM approach for coupling ROM and FOM descriptions, we introduce an array of interface modeling paradigms as depicted in Fig~\ref{fig:interface} and described next. 

The \emph{Direct Prolongation Interface (DPI)} approach utilizes standard projection based ROMs, where the system's state at the interface is obtained by the reconstruction of a Galerkin projection ROM solution. However, traditional Galerkin ROM often yields an inaccurate solution in case of systems with strong nonlinearity. Therefore, we utilize machine learning to correct and augment ROM solution in a hybrid framework, as follows. 

In the \emph{Prolongation followed by machine learning Correction Interface (PCI)} methodology, an LSTM is used to rectify the field reconstruction from Galerkin ROM at the interface by learning the correction in the higher dimensional space. Although this seems to be a straightforward implementation, it might amount to learning a high dimensional correction vector, especially for two- and three-dimensional domains. 

To mitigate the potential computational challenges dealing with excessively large input/output vectors, a \emph{machine learning Correction followed by Prolongation Interface (CPI)} approach can be employed to provide a closure effect to remedy the instabilities and inaccuracies of Galerkin ROM due to modal truncation.

For CPI, the LSTM learns the correction terms in ROM space, defined by the number of modes in ROM approximation. As a result, the reconstruction quality will eventually be limited by the Galerkin ROM dimension. Therefore, the \emph{Uplifted Prolongation Interface (UPI)} framework not only corrects the Galerkin ROM solution, but also expands the ROM subspace to enhance the reconstruction quality. Our primary motivation in this paper is to describe and test these four interface modeling approaches to tackle ROM/FOM coupling problems and show how we can elucidate these multifidelity mechanisms within the HAM framework.


\section{ROM-FOM Coupling Framework}
\textcolor{rev}{In order to demonstrate the performance of the introduced HAM approaches for ROM-FOM coupling, we consider a coupled system as follows,} 
\begin{align}
    \dfrac{\partial u}{\partial t} = f_1(u;\mu_1) + g_1(u,v;\mu_1,\mu_2), \label{eq:u}\\
    \dfrac{\partial v}{\partial t} = f_2(v;\mu_2) + g_2(u,v;\mu_1,\mu_2),
\end{align}
\textcolor{rev}{where $u$ and $v$ are the coupled variables and $g_1$ and $g_2$ define this coupling, while $\mu_1$ and $\mu_2$ denote the set of system's parameters. We highlight that the coupled variables might represent the state variables at different regions of the domain (e.g., multi-component systems), different physics (e.g., fluid-structure interactions) and/or different scales within the same domain (e.g., multiscale systems). We suppose that the dynamics of $u$ can be approximated by a reduced order model (ROM) while a full order model resolves $v$ and both solvers need to communicate information to satisfy the coupling. We begin by describing the derivation of a reduced order model of $u$ via Galerkin projection equipped with proper orthogonal decomposition (POD) for basis construction. Then, we formulate the coupling between ROM and FOM solvers.}

\subsection{Reduced order model}
Introducing a spatial discretization to Eq.~(\ref{eq:u}), it can be rewritten in a semi-discrete continuous-time as follows, 
\begin{equation}
    \dfrac{\mathrm{d} \mathbf{u}}{\mathrm{d} t} = \mathcal{F}(\mathbf{u}, \mathbf{v}; \boldsymbol{\mu}) = \mathcal{L}_1 \mathbf{u} + \mathcal{L}_2 \mathbf{v} + \mathcal{N}(\mathbf{u},\mathbf{v}), \label{eq:udis}
\end{equation}
where the boldfaced symbols represent the arrangement of discretized variables in 1D vector (e.g., $\mathbf{u} \in \mathbb{R}^{n_1}$ and $\mathbf{v} \in \mathbb{R}^{n_2}$, where $n_i$ denotes the spatial resolution), $\boldsymbol{\mu} \in \mathbb{R}^p$ defines the system's parameters, and $\mathcal{F}: \mathbb{R}^{n_1} \times \mathbb{R}^{n_2} \times \mathbb{R}^p \to \mathbb{R}^n$ is a deterministic operator with linear and nonlinear components $\mathcal{L}$, and $\mathcal{N}$, respectively. These operators depends on the numerical scheme adopted for spatial discretization.

We exploit the advances and developments of ROM techniques to build surrogate models to economically resolve portions of domain and/or physics. The ROM solution can thus be used to infer the flow conditions at the interface so that a FOM solver can be efficiently employed for the sub-domains of interest. The standard Galerkin ansatz is applied for the dynamics of $\mathbf{u}$ as 
\begin{equation}
    \mathbf{u}(t) \approx \Phi \boldsymbol{\alpha}(t), \label{eq:POD}
\end{equation}
where the columns of matrix $\Phi = [\phi_1, \phi_2, \dots, \phi_r] \in \mathbb{R}^{n_1\times r}$ form the orthonormal bases of a reduced subspace with an intrinsic dimension of $r$, and $\boldsymbol{\alpha}$ defines the projection coordinates associated with $\Phi$. Usually, the basis functions $\phi$ are constructed to capture the dominant modes or underlying structures of the flow. Proper orthogonal decomposition (POD) is one popular technique to systematically construct $\Phi$ such that the solution manifold preserves as much variance as possible when projected onto the subspace spanned by $\Phi$ \cite{sirovich1987turbulence,berkooz1993proper,holmes2012turbulence}. By substituting this approximation into Eq.~(\ref{eq:udis}) and performing the inner product with $\Phi$, we get the following,
\begin{equation}
 \dfrac{\mathrm{d} \boldsymbol{\alpha}}{\mathrm{d} t} = \Phi^T \mathcal{L}_1 \Phi \boldsymbol{\alpha} + \Phi^T \mathcal{L}_2 \mathbf{v} + \Phi^T \mathcal{N}(\Phi \boldsymbol{\alpha}, \mathbf{v}). \label{eq:ROM1}   
\end{equation}
The first coefficient ($\Phi^T \mathcal{L}_1 \Phi$) can be precomputed, so the computational cost for evaluating the linear term depends on $r$. However, in general, the evaluation of the third term on the right hand side (nonlinear term) depends on the FOM dimension $n$. Fortunately, most fluid flow systems are characterized by quadratic nonlinear operator, which allows the reduction of Eq.~(\ref{eq:ROM1}) into 
\begin{equation} 
\dfrac{\mathrm{d} \boldsymbol{\alpha}}{\mathrm{d} t} = \mathbf{\mathfrak{L}} \boldsymbol{\alpha}  +  \boldsymbol{\alpha}^T \mathbf{\mathfrak{N}} \boldsymbol{\alpha} + \mathbf{\mathfrak{C}} , \label{eq:ROM}
\end{equation}
where $\mathbf{\mathfrak{L}}$ is an $(r\times r)$ matrix and $\mathbf{\mathfrak{N}}$ is an $(r\times r \times r)$ tensor representing the model coefficients while $\mathbf{\mathfrak{C}}$ defines the contribution of $\mathbf{v}$ into the ROM of $\mathbf{u}$. We will see that the computation of $\mathbf{\mathfrak{C}}$ may either be computed offline during ROM construction or as part of the online FOM solver of $\mathbf{v}$ with negligible computational overhead. Thus, the floating point operation (flop) count to evaluate the right hand side of the ROM (i.e., Eq.~(\ref{eq:ROM})) is often $O(r^3)$.

\begin{figure}[ht]
\centering
\includegraphics[width=0.95\linewidth]{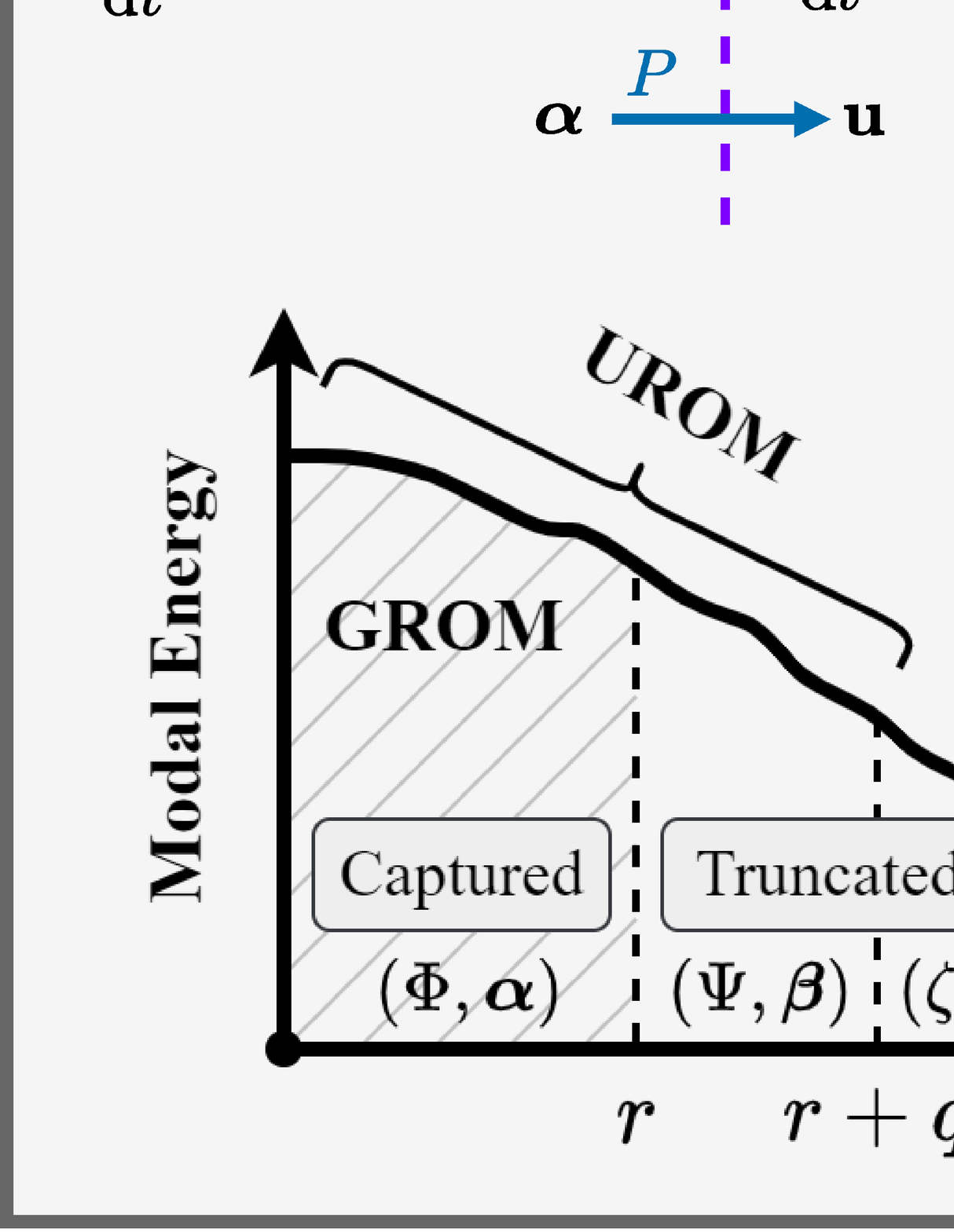}
\caption{Schematic illustration of the methodologies introduced to utilize ROM to economically provide sound interface conditions in a multifidelity domain decomposition problems. Galerkin ROM yields inaccurate predictions (represented by rough curve), and direct prolongation of these results might be not efficient. PCI adds a correction effect to the prolonged solution in FOM space. Instead, CPI and UPI introduce the correction at ROM level before prolongation. UPI adds an extra superresolution effect to augment solution quality.}
\label{fig:def}
\end{figure}

In the following, we formulate the four methodologies outlined in Fig~\ref{fig:interface} to match the ROM solution for $\boldsymbol{\alpha}$ with the FOM solution at the interface. For all cases, a ROM representation is adopted for $u$, which can be economically solved to compute an estimate of the interface flow condition to feed the FOM solver of $v$. For example, in multi-component systems like that depicted in Fig.~\ref{fig:def}, the ROM solution at the interface is regarded as a boundary condition for the FOM.

\begin{enumerate}
    \item \emph{DPI: Direct Prolongation Interface.} The objective of the DPI approach is to recover the flow variables at the interface from the ROM solution (i.e., the time integration of Eq.~(\ref{eq:ROM})). In other words, we seek to learn a mapping $\mathcal{G}_1:\mathbb{R}^r \to \mathbb{R}^d$, that minimizes $\|u^{(i)} - \mathcal{G}_1(\boldsymbol{\alpha})\|$ where $u^{(i)}$ represent the portion of information at the interface that is shared from the ROM to the FOM solver, with $d$ being the dimension of the interface. For multi-component systems, this interface can be a single point (e.g., for 1D systems), a line (e.g., for 2D systems), or a surface (e.g., for 3D systems). Indeed, this prolongation map naturally results from the Galerkin ansatz, and can be written as 
    \begin{equation}
    \mathcal{G}_1(\boldsymbol{\alpha}) = \boldsymbol{\Theta} \boldsymbol{\alpha}, \label{eq:DPI}
    \end{equation}
    where $\boldsymbol{\Theta}$ represents the portion of the basis $\Phi$ that is computed at the interface location. 
    
    Since the ROM approximation is built upon the assumption of representing the flow within a low order subspace, the approximation given by Eq.~(\ref{eq:POD}) basically introduces a projection error. This error can be significant for complex systems, where the flow dynamics are characterized by a wide spectrum while only few modes are considered to minimize the computational burden of solving the ROM. Moreover, the nonlinear interactions as well as the modal truncation coupled with the Galerkin projection methodology usually cause Eq.~(\ref{eq:ROM}) to yield erroneous predictions of the coefficients $\boldsymbol{\alpha}(t)$. Therefore, the solution from the DPI approach is potentially inaccurate \cite{ahmed2020long}. Consequently, the reconstruction $\mathcal{G}_1(\boldsymbol{\alpha})$ is no longer optimal and a correction needs to be introduced. 
    


   
    \item \emph{PCI: Prolongation followed by Correction Interface.} The PCI framework aims to correct the mapping $\mathcal{G}_1$ to yield more accurate interface condition. To do so, we utilize a long short-term memory (LSTM) neural network to learn a mapping $\mathcal{G}_2:\mathbb{R}^i \to \mathbb{R}^i$ such that
    \begin{equation}
    \mathcal{G}_2\left(\mathcal{G}_1(\boldsymbol{\alpha})\right) = u^{(i)} - \mathcal{G}_1(\boldsymbol{\alpha}).
    \end{equation}
    In other words, LSTM is fed with a predictor of $u^{(i)}$ obtained by DPI and approximates the deviation of this value from the true state variables at the interface. Hence, this deviation estimate can be added as a correction term in a predictor-corrector fashion. In PCI, both inputs and outputs of the LSTM lie in the FOM space and thus the LSTM map can be considered as nudging scheme from the ROM prolongation $\mathcal{G}_1$ to the FOM solution \cite{pawar2020long}.

    We highlight that the PCI approach can be feasible for one-dimensional (1D) problems (where the interface can be just a single point). However, for higher dimensional systems, the sizes of input and output vectors grow rapidly (unless a too coarse mesh is adopted). For such cases, PCI becomes prohibitive, and the learning and correction should be performed in a reduced latent space instead.


    \item \emph{CPI: Correction followed by Prolongation Interface.} The CPI methodology works by introducing the correction in the latent subspace, rather than the FOM space. This is especially crucial for 2D and 3D configurations. In particular, the CPI aims at curing the deviation in modal coefficients predicted from solving the Galerkin ROM equations, known as closure error. Due to the modal cut-off in ROM approximation, Eq.~(\ref{eq:ROM}) does not necessarily capture the true projected trajectory of $\boldsymbol{\alpha}(t)$. Therefore, we introduce an LSTM mapping $\mathcal{G}_3: \mathbb{R}^r \to \mathbb{R}^r$ to provide a closure effect to adjust the Galerkin ROM trajectory. Specifically, the LSTM for CPI takes the values of modal coefficients acquired from integrating Eq.~(\ref{eq:ROM}) in time and predicts the discrepancy between these values and their optimal values. Those are defined by the true projection (TP) of the FOM solution onto the basis functions as follows,
    \begin{equation}
        \boldsymbol{\alpha}^{\text{TP}} = \Phi^T \mathbf{u}.
    \end{equation}
    Therefore, the CPI contribution can be written as
    \begin{equation}
        \mathcal{G}_3(\boldsymbol{\alpha}) =  \Phi^T \mathbf{u} - \boldsymbol{\alpha}.
    \end{equation}
    We highlight here that the size of the input and output vectors is $O(r)$, independent of the FOM resolution, which offers a potential flexibility dealing with 2D and 3D problems. Once the modal coefficients are corrected, they are prolonged from the ROM space to the FOM space using the reconstruction map $\mathcal{G}_1$. For all results, we also show the results obtained from the true projection of FOM solution onto the ROM subspace at the interface as,
    \begin{equation}
        \mathbf{u}^{\text{TP}} = \mathcal{G}_1(\boldsymbol{\alpha}^{\text{TP}}).
    \end{equation}
    We highlight that $u^{\text{TP}}$ represents the best approximation of the true flow field that can be achieved using a linear subspace with an intrinsic dimension of $r$.

    
    \item \emph{UPI: Uplifted Prolongation Interface.} Although the CPI methodology cures the closure error and provides a stabilized solution, it does not address projection error. Unless a large number of modes are resolved, the projection error can be significant, especially for problems with discontinuities and shocks. To deal with those situations, an uplifting ROM has been proposed \cite{ahmed2020long}, where both closure and projection errors are taken care of. For closure, similar to CPI, the Galerkin ROM predictions are tuned to match their true projection values. In addition, following Galerkin ROM solution, the ROM subspace is expanded to capture some of the smaller scales missing in the initial subspace as follows,
    \begin{equation}
        \mathbf{u} \approx \Phi \boldsymbol{\alpha} + \Psi \boldsymbol{\beta},
    \end{equation}
    where the columns of $\Psi = [\psi_1, \psi_2, \dots, \psi_q]$ form orthonormal basis functions for a $q$-dimensional subspace complementing that spanned by $\Phi$ and $\boldsymbol{\beta}$ defines the corresponding projection coordinates. Similar to $\Phi$, $\Psi$ can be computed through the POD algorithm. Note that $\Phi$ and $\Psi$ are orthogonal to each others (i.e., $\Phi^T \Psi = \Psi^T \Phi = \mathbf{0}$). Indeed, $\Psi$ represents the next $q$ basis functions generated by POD after the first $r$ being dedicated to $\Phi$. Those are also constructed a priori during an offline stage using the collected set of snapshot data. We highlight that the Galerkin ROM equations only solve for $\boldsymbol{\alpha}$ to keep the computational cost as low as possible. 
    
    Therefore, a complementary model for $\boldsymbol{\beta}$ has to be constructed so that the uplifting approach can be employed. To accomplish this, a mapping from the first $r$ modal coefficients to the next $q$ modes is assumed to exist. Nonlinear Galerkin projection has been pursued to express this mapping as $\boldsymbol{\beta} = \mathcal{H}(\boldsymbol{\alpha})$, but it has been found challenging for most systems \cite{wan2018data}. Instead, we exploit the LSTM learning capabilities to infer this map from data. This uplifting approach enhances the quality of prolonged solution by providing a superresolution effect. In particular, the UPI architecture is trained to read the Galerkin ROM prediction for the first $r$ modal coefficients as input, and return the true coefficients of the first $r+q$ modes. Thus, it provides a closure effect for the first $r$ modes and a superresolution effect for the next $q$ modes, simultaneously in a single network as follows,
    \begin{align}
        \mathcal{G}_4&: \mathbb{R}^r \to \mathbb{R}^{r+q}\\
        \mathcal{G}_4(\boldsymbol{\alpha}) &= \begin{bmatrix} \boldsymbol{\alpha}^{\text{TP}} \\  \boldsymbol{\beta}^{\text{TP}}\end{bmatrix}.
    \end{align}
    We note here that the first $r+q$ spatial modes have to be built and stored beforehand, which introduces slightly more storage costs. For the present study, we explore the specific case where $q=r$, but a generalization is straightforward.
\end{enumerate}

The ROM-FOM coupling philosophy as well as the introduced interface learning approaches are summarized in the cartoon shown in Fig~\ref{fig:def}. These methodologies are also applicable to a wide range of computational problems with multifidelity domain decomposition. The depicted system is assumed to be fully characterized by three mutually orthogonal sets of basis functions, namely $\Phi$, $\Psi$, and $\zeta$ as below
\begin{equation}
    \mathbf{u} = \Phi \boldsymbol{\alpha} + \Psi \boldsymbol{\beta} + \zeta \boldsymbol{\gamma},
\end{equation}
where $\boldsymbol{\alpha}$, $\boldsymbol{\beta}$, and $\boldsymbol{\gamma}$ are the corresponding projection coordinates. We also suppose that Galerkin ROM resolves the $\Phi$ set of modes (i.e., truncating the contributions of $\Psi$, and $\zeta$). We reiterate here that the Galerkin ROM yields inaccurate solution (sketched by the noisy (rough) curve of predicted $\boldsymbol{\alpha}$). Consequently, the quality of the direct prolongation mapping is compromised. The PCI aims at correcting the reconstructed solution at the interface. Even though the PCI technique acts only on a small portion of the domain (i.e., the interface), it might amount to excessively large input and output sizes. 

On the other hand, CPI treats the Galerkin ROM deficiencies at the ROM level. In particular, it introduces a closure effect to better predict $\boldsymbol{\alpha}$. This closure simply compensates the effects of truncated modes (i.e., $\Psi$ and $\zeta$) onto the dynamics of $\Phi$. This yields a better estimate of $\boldsymbol{\alpha}$, as illustrated by the smooth curve in Fig~\ref{fig:def}. We note that in CPI, the effects of $\Psi$ and $\zeta$ are only considered to improve the prediction of $\boldsymbol{\alpha}$. However, their contributions to the solution manifold reconstruction are not included, resulting in a substantial reconstruction error (projection error). To deal with this caveat, UROM seeks to add a superresolution enhancement by incorporating the $\Psi$ set of basis into the reconstruction step by learning the dynamics of the corresponding $\boldsymbol{\beta}$ coordinates. This is represented by a higher resolution (denser) reconstruction in UPI case, compared to CPI, and DPI. Note that the PCI is still showing the highest resolution (the densest reconstruction) as it nudges the prediction at the interface to its FOM counterpart (i.e., including all $\Phi$, $\Psi$, and $\zeta$).

\section{Demonstrations}
\textcolor{rev}{We demonstrate the ROM-FOM coupling methodologies using two examples of varying complexities. In the first one, we describe a fluid flow scenario over a bizonal domain with heterogeneous physical properties using the one-dimensional Burgers problem. For this case, we shall see that the interface between different sub-domains is defined by a single point (i.e., $d=1$). Second, we consider the Marsigli flow problem represented by the two-dimensional Boussinesq equations to demonstrate the ROM-FOM coupling for multiphysics systems. In particular, a ROM solver is devoted for the mass and momentum transport equations while a FOM is reserved for the energy transport.}

\textcolor{rev}{\subsection{The one-dimensional Burgers problem}}
In order to represent a zonal multifidelity simulation, we consider the following one dimensional (1D) viscous Burgers problem,
\begin{align}
    \dfrac{\partial u}{\partial t} &+ u \dfrac{\partial u}{\partial x} = \dfrac{\partial}{\partial x} \left(\nu \dfrac{\partial u}{\partial x}\right) - \gamma u, \label{eq:brg} \\
    (\nu, \gamma) &= \begin{cases}
            (10^{-2},0) \quad \text{for } 0\le x \le x_b\\
            (10^{-4},1) \quad \text{for } x_b < x \le 1,
            \end{cases}
\end{align}
where $x_b$ is the spatial location of the interface defining the physical heterogeneity. We highlight that Eq.~(\ref{eq:brg}) includes the $\dfrac{\partial}{\partial x} \left(\nu \dfrac{\partial u}{\partial x}\right)$ term instead of the commonly used $\nu \dfrac{\partial^2 u}{\partial x^2}$ term to permit the spatial variation of $\nu$. For the given setup, the stiffness and physical properties in the left part dictates higher spatial as well as temporal resolutions than those required for the right partition. If we opt to a global unified (unique fidelity) solver over the whole domain, the quality of the solution will be dominated by stiffness of the left zone. In other words, a smaller time step would be required to satisfy numerical stability when using an explicit temporal integration scheme. Specifically, assuming that we utilize the forward in time and central in space finite difference scheme (FTCS) to solve Eq.~(\ref{eq:brg}) with a spatial grid resolution of $4096$, a time step of approximately $2.5\times10^{-6}$ will be required for the left part of domain, while a time step of $2.5\times10^{-4}$ would be sufficient if we were able to only resolve the right part. Therefore, domain decomposition approaches might be adopted to segregate partitions with varying numerical requirements. Despite the effectiveness and efficiency of these approaches, idle delays can arise in order to accommodate information transfer from the left zone to the right zone through their common interface. Instead, low-fidelity proxy models can be utilized to avoid such lags by approximating the \emph{effective} dynamics of stiff regions and providing sound interface boundary conditions to the rest of the computational domain. 

The discretized domain is divided into a left zone with $\mathbf{u}^{L} \in \mathbb{R}^{n_1}$ for $x\in [0,x_b]$ and a right zone with $\mathbf{u}^{R} \in \mathbb{R}^{n_2}$ for $x\in [x_b,1]$, where $n_1 + n_2 = n+1$. We build a reduced order model for the left sub-domain as follows,
\begin{align} 
\mathbf{u}^{L}(t) &= \Phi \boldsymbol{\alpha}(t), \label{eq:uPOD} \\
\dfrac{\mathrm{d} \boldsymbol{\alpha}}{\mathrm{d} t} &= \mathbf{\mathfrak{L}} \boldsymbol{\alpha}  +  \boldsymbol{\alpha}^T \mathbf{\mathfrak{N}} \boldsymbol{\alpha}, \label{eq:uROM}
\end{align}
where $\mathbf{\mathfrak{L}}$ and $\mathbf{\mathfrak{N}}$ represent the tensorial ROM coefficients which can be precomputed during the offline stage as,
\begin{align}
    \mathfrak{L}_{i,k} & = \big\langle \dfrac{\partial}{\partial x} \left(\nu \dfrac{\partial u}{\partial x}\right) - \gamma \phi_i;  \phi_k \big\rangle, \\
    \mathfrak{N}_{i,j,k} &= \big\langle -\phi_i \dfrac{\partial \phi_j}{\partial x};  \phi_k \big\rangle,
\end{align}
where the angle parentheses denote the inner product (i.e., $\langle \mathbf{a} ; \mathbf{b} \rangle = \mathbf{a}^T \mathbf{b}$).

For data generation, we solve the full order model representing the 1D Burgers equation (Eq.~(\ref{eq:brg})) over the entire domain using a spatial mesh resolution of $4096$ and time step of $2.5\times10^{-6}$. For initial condition, we consider a square-like wave defined as,
\begin{equation}
    u(x,0) = 0.5 - 0.5 \tanh{\dfrac{x-x_b}{\epsilon}},
\end{equation}
where a value of $x_b = 0.75$ is considered as the location of the interface and $\epsilon=0.005$ is used to define the sharpness of the shock at $x_b$. Dirichlet boundary conditions are assumed at both boundaries (i.e., $u(0,t)=u(1,t)=0$). We compute the evolution of the velocity field for $t\in [0,2]$, and collect snapshots every $100$ time steps. That is snapshots are collected every $2.5\times 10^{-4}$ time units (working with normalized variables). 

For ROM construction, we consider the truncated solution snapshots for the left part of the domain (i.e., $0\le x\le 0.75$) for $t \in [0,1]$. For POD basis generation, we use only $200$ snapshots distributed evenly from $t=0$ to $t=1$, to reduce the computational cost of solving the corresponding eigenvalue problem. Once ROM is constructed, it is integrated in time with a time step of $2.5\times10^{-4}$ to match the time step in the right part of the domain (to be handled via a FOM solver). During the deployment phase of the coupled system, the ROM feeds the FOM solver with the boundary condition at $x_b$. On the other hand, the effects of the interface on the ROM dynamics are considered during the offline stage of data generation and basis construction.

\textcolor{rev}{We utilize labelled data at the time interval of $t\in [0,1]$ for the training and validation of the LSTM neural networks. In particular, four fifths of the collected data set are randomly selected for training while the remaining one fifth is reserved for validation purposes (to avoid overfitting). We highlight that numerical experiments reveal that the performance in our case is not significantly sensitive to the neural network hyperparameters. Thus, we manually tune the hyperparameters during the training and validation stages by repeating the previous procedure multiple times while varying the seeds for the random number generator and  using averaged loss values for assessment. We adopt an LSTM architecture of 2 layers with 20 cells in each layer. In the meantime, we note that automated hyperparameter search tools can be utilized for optimized architectures. Testing of the proposed schemes is performed for $t \in [0,2]$, corresponding to a temporal extrapolation behavior with respect to the training interval.} 

\begin{figure}[ht]
\centering
\includegraphics[width=0.99\linewidth]{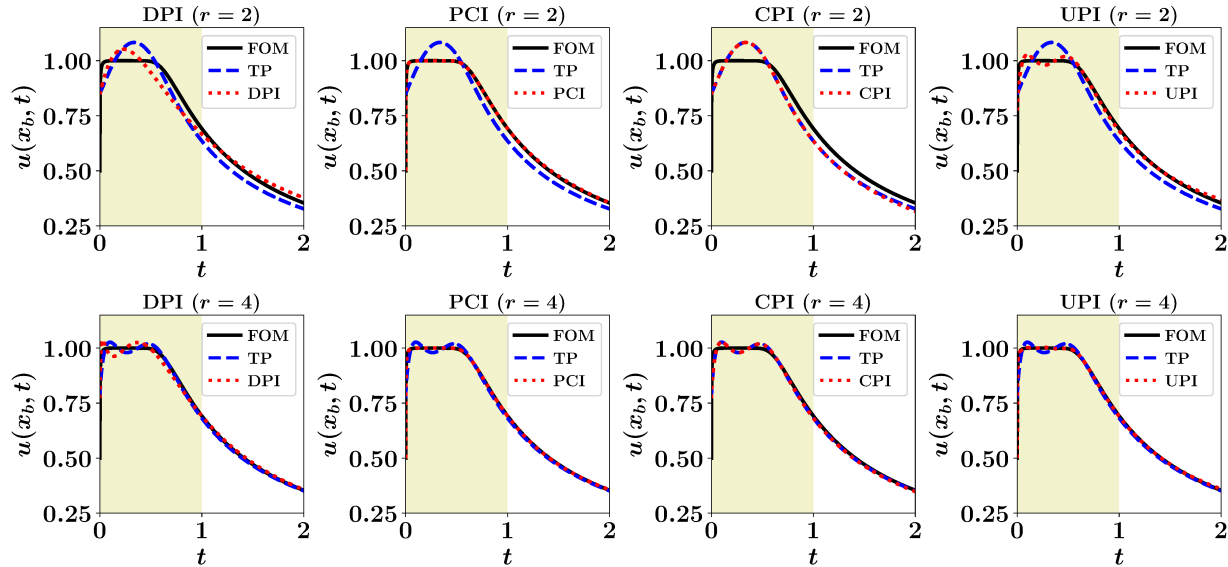}
\caption{Velocity at the interface obtained by considering ROM in the left part of the domain,  with $r=2$ (top) and $r=4$ (bottom). FOM solution corresponds to solving the governing equation over the entire domain, while TP denotes the projection of the FOM solution in the left zone onto the corresponding POD subspace.}
\label{fig:r24}
\end{figure}

\textcolor{rev}{Since the coupling between ROM and FOM is represented by the physical interface at $x_b$, we notice that the $\boldsymbol{\Theta}$ in the DPI map (i.e., $\mathcal{G}_1(\boldsymbol{\alpha}) =  \boldsymbol{\alpha}$) is simply the last row of the matrix $\boldsymbol{\Phi}$.} In Fig~\ref{fig:r24}, we plot the velocity at the interface obtained from adopting ROM for the left part of the domain, and corrected with machine learning architectures as described before for $r=2$ and $r=4$. We note that FOM solution corresponds to the velocity at the interface obtained by applying a FOM solver all over the domain using a time step of $2.5 \times 10^{-6}$, while the true projection (TP) represents the projection of the truncated velocity field in the left zone onto the POD subspace. The shaded area in Fig~\ref{fig:r24} stands for the time interval utilized for POD basis generation, ROM formulation, and LSTMs training. 

It can be seen that DPI results, especially for $r=2$, are not very accurate due to the mutual effects of modal truncation and model nonlinearity in Galerkin ROM. On the other hand, the PCI solution gives almost perfect match with FOM. As the PCI approach nudges the prolonged ROM solution to its FOM counterpart, it gives even higher accuracy than TP. That is TP is limited by the maximum quality that can be obtained using a rank-$r$ approximation. For CPI, since the LSTM introduces a closure effect, it steers the ROM results to match the TP solution. Finally, the UPI recovers some of the smaller scales (truncated modes) so it yields better reconstruction than TP since it spans a larger subspace. For $r=2$, UPI uplifts the solution to a rank-4 approximation, while for $r=4$, it is uplifted to a rank-8 approximation.

For quantitative assessment, we document the $\ell_2$ norm of the deviation of the predicted velocity at the interface compared to the FOM solution in Table~\ref{table:norm}. Results are reported for $r\in \{2,4,8\}$. We see that the error in the CPI case almost matches that of TP, while PCI gives the highest accuracy since it is trained to learn the correction with respect to the FOM solution. Also, CPI results are significantly close to TP, illustrating the closure effect introduced by CPI to account for the effect of modal truncation on ROM dynamics. Another interesting observation is that UPI quality at a given value of $r$ is equivalent to TP with twice that value. This indicates that UPI is able to give a superresolution effect up to $2r$ (since we select $q=r$). Also, we notice that at $r=2$, DPI yields lower $\ell_2$ norm than TP. This is because TP solution is obtained by the projection of the FOM solution onto the POD subspace generated using data at $t\in[0,1]$. So, this subspace is optimal only for $t\in[0,1]$, while testing is performed up to $t=2$. Therefore, TP solution no longer represents the best rank-$r$ approximation beyond $t=1$.

\begin{table}[htbp!]
\caption{$\ell_2$ norm for the deviation of the velocity at the interface with respect to its FOM value for $t\in [0,2]$.}
\centering
\begin{tabular}{p{0.08\textwidth} p{0.08\textwidth} p{0.08\textwidth} p{0.08\textwidth} }  
\hline
Setup  & $r=2$ & $r=4$ & $r=8$ \\
\hline 
TP   & $4.56$ & $1.25$ & $0.21$\\ 
DPI  & $3.67$ & $1.46$ & $0.44$\\ 
PCI  & $0.23$ & $0.07$ & $0.03$\\ 
CPI  & $4.57$ & $1.27$ & $0.23$\\ 
UPI  & $1.28$ & $0.24$ & $0.11$\\ 
\hline
\end{tabular}
\label{table:norm}
\end{table}

\begin{figure}[ht]
\centering
\includegraphics[width=0.95\linewidth]{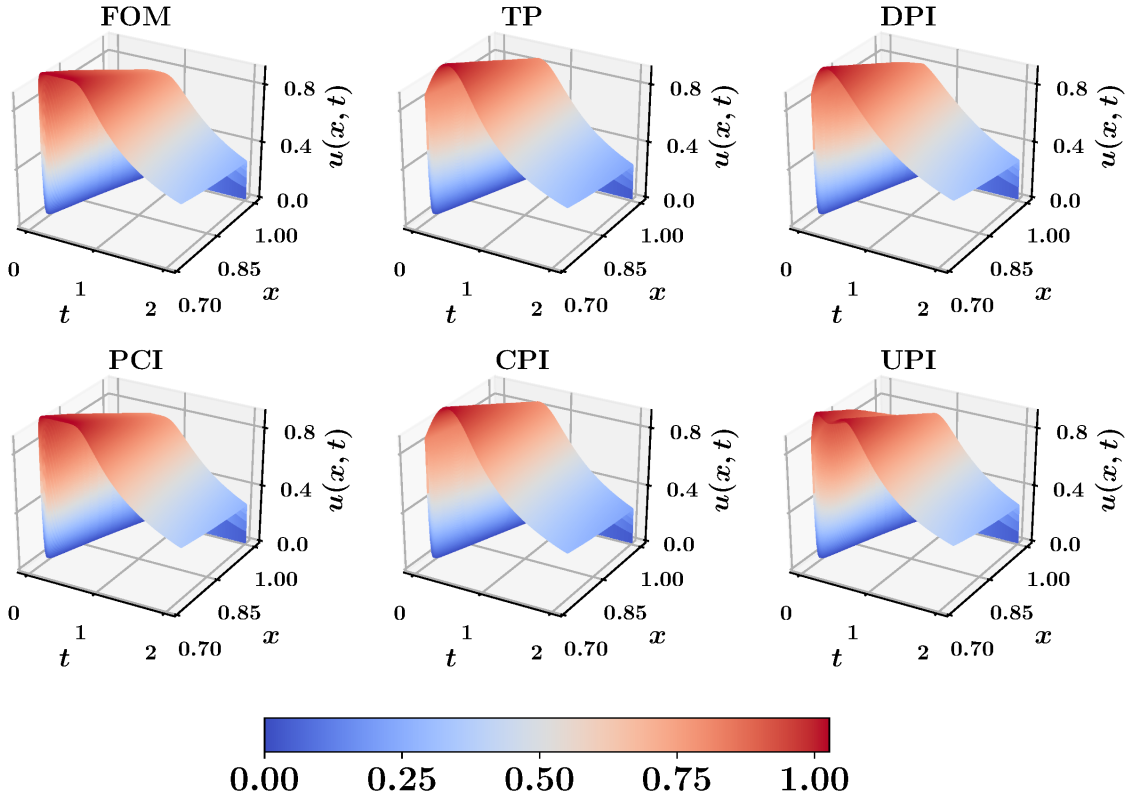}
\caption{Spatio-temporal velocity profile obtained from applying high fidelity (FOM) solver onto the right subdomain ($0.75 \le x \le 1$), fed with interface boundary from a low-fidelity (ROM) solution with $r=2$.}
\label{fig:r2surf}
\end{figure}

Finally, we investigate the coupling efficiency by solving the right part (i.e., $0.75 \le x \le 1$) using a high fidelity FOM solver applied only onto this subdomain. This is fed with a boundary condition $u(0.75,t)$ from the low-fidelity interface learning approaches described before. Fig~\ref{fig:r2surf} shows the spatiotemporal velocity profile with $r=2$, compared to the FOM predictions. Again, we observe that the solution with PCI boundary is similar to this FOM solution. Also, CPI matches the TP results, but they both smooth-out the surface peak because of the low rank limitations. \textcolor{rev}{Although it seems that the accuracy of UPI cannot exceed that of CPI with $r+q$ assuming optimal performance for both CPI and UPI, there is a computational side in this comparison. The CPI with $r+q$ modes implies the solution of a Galerkin ROM with a dimension of $r+q$, while the UPI requires the solution of a Galerkin ROM with $r$. We have seen that for fluid flows with quadratic nonlinearity, the computational cost of solving a Galerkin ROM scales cubically with the number of modes. Thus, the implementation of CPI with $r+q$ with $q=r$ is about 8 times more costly than UPI with $r+q$. At this point, we highlight that the selection of $r$ and $q$ is highly dependent on the problem in hand, the corresponding decay of POD eigenvalues, and the level of accuracy sought. A compromise between computational cost of solving a Galerkin ROM with $r$ and the corresponding stability, the amount of information captured by $r$ and $r+q$ modes, and reliability of UPI with $r+q$ is always in place.}

\textcolor{rev}{\subsection{The two-dimensional Boussinesq problem}} 
\textcolor{rev}{Boussinesq equations represent a simple approach for modeling geophysical waves such as oceanic and atmospheric circulations induced by temperature differences \cite{majda2003introduction} as well as other situations, like isothermal flows with density stratification. In the Boussinesq approximation, variations of all fluid properties other than the density are neglected completely. Moreover, the density dependence is ignored in all terms except for gravitational force (giving rise to buoyancy effects). As a result, the continuity equation can be adopted in its constant density form, and the momentum equation can be simplified significantly. The dimensionless form of the 2D incompressible Boussinesq equations in vorticity-streamfunction formulation can be represented by the following two coupled transport equations \cite{liu2003fourth,nicolas20052d},
\begin{align}
\dfrac{\partial \omega}{\partial t} + J(\omega,\psi) &= \dfrac{1}{\text{Re}}\nabla^2 \omega + \text{Ri} \dfrac{\partial \theta}{\partial x}, \label{eq:Bouss1} \\
\dfrac{\partial \theta}{\partial t} + J(\theta,\psi) &= \dfrac{1}{\text{Re} \text{Pr}} \nabla^2 \theta, \label{eq:Bouss2}
\end{align}
where $\omega$, $\psi$ and $\theta$ denote the vorticity, streamfunction and temperature fields, respectively. In Boussinesq flow systems, there are  three leading physical mechanisms, namely viscosity, conductivity, and buoyancy, giving rise to three dimensionless numbers; Reynolds number $\text{Re}$, Richardson number $\text{Ri}$, and Prandtl number $\text{Pr}$.}

\textcolor{rev}{We utilize the 2D Boussinesq equation to illustrate the ROM-FOM coupling for multiphysics situations. In particular, we suppose that we are more interested in the temperature field predictions. Thus, we dedicate a FOM solver for Eq.~(\ref{eq:Bouss2}). However, we see that the solution of this equation requires evaluating the streamfunction field at each time step. The kinematic relationship between vorticity and streamfunction is given by the following Poisson equation,
\begin{equation}\label{eq:Poisson}
\nabla^2 \psi = -\omega,
\end{equation}
which implies that the streamfunction is not a prognostic variable, and can be computed from the vorticity field at each timestep. In typical simulations, the solution of Eq.~(\ref{eq:Poisson}) consumes significant amount of time and computational resources and is considered the bottleneck for most incompressible flow solvers. Therefore, we can consider a ROM solver for the voriticity dynamics as follows,
\begin{align}
        \omega(x,y,t) &= \bar{\omega}(x,y) + \sum_{k=1}^r \alpha_k(t) \phi_k^{\omega}(x,y),
\end{align}
where $\bar{\omega}$ denotes the time-averaged vorticity field and the POD is performed on the fluctuating part of $\omega$. We note that Eq.~(\ref{eq:Poisson}) allows us to assume the same modal coefficients $\alpha_k(t)$ for both $\omega$ and $\psi$ as follows,
\begin{equation} \label{eq:sPOD}
    \psi(x,y,t) = \bar{\psi}(x,y) + \sum_{k=1}^r \alpha_k(t) \phi_k^{\psi}(x,y),
\end{equation}
where the time-averaged streamfunction field and the corresponding basis functions can be computed from the following relations,
\begin{align}
    \nabla^2 \bar{\psi}(x,y) &= -\bar{\omega}(x,y), \\
    \nabla^2 \phi_k^{\psi}(x,y) &= -\phi_k^{\omega}(x,y).
\end{align}}
\textcolor{rev}{Thus, the Galerkin ROM of Eq.~(\ref{eq:Bouss1}) can be written as
\begin{align} \label{eq:wROM}
    \dfrac{\text{d}\alpha_k}{\text{d}t} &=  \mathfrak{B}_k + \sum_{i=1}^R \mathfrak{L}^{(\omega,\psi)}_{i,k} \alpha_i + \sum_{i=1}^R \mathfrak{L}^{(\omega,\theta)}_{i,k} \beta_i + \sum_{i=1}^R \sum_{j=1}^R \mathfrak{N}^{(\omega,\psi)}_{i,j,k} \alpha_i \alpha_j,
\end{align}
where $\beta$ represent the projection of the temperature fields onto the reduced subspace defined as \begin{equation} \label{eq:beta}
    \beta_k(t) = \big \langle \theta(x,y,t_n) - \bar{\theta}(x,y) ; \phi_k^{\theta}(x,y) \big \rangle,
\end{equation} 
where the $\bar{\theta}$ denotes the time-averaged field of $\theta$ and $\boldsymbol{\Phi}^{\theta}$ are the corresponding orthonormal POD modes. The predetermined coefficients in Eq.~(\ref{eq:wROM}) are calculated as follows,
\begin{align}
    \mathfrak{B}_k &= \big\langle -J(\bar{\omega},\bar{\psi}) + \dfrac{1}{\text{Re}} \nabla^2 \bar{\omega} + \text{Ri} \dfrac{\partial \bar{\theta}}{\partial x}; \phi_k^{\omega} \big\rangle, \\
    \mathfrak{L}^{(\omega,\psi)}_{i,k} &= \big\langle \dfrac{1}{\text{Re}} \nabla^2 \phi_i^{\omega} - J(\phi_i^{\omega},\bar{\psi}) - J(\bar{\omega},\phi_i^{\psi}) ; \phi_k^{\omega} \big\rangle, \\
    \mathfrak{L}^{(\omega,\theta)}_{i,k} &= \big\langle \text{Ri} \dfrac{\phi_i^{\theta}}{\partial x} ; \phi_k^{\omega} \big\rangle, \\
    \mathfrak{N}^{(\omega,\psi)}_{i,j,k} & = \big\langle -J(\phi_i^{\omega},\phi_j^{\psi}) ; \phi_k^{\omega} \big\rangle.
\end{align} }

\textcolor{rev}{We notice that the ROM defined by Eq.~(\ref{eq:wROM}) equipped with Eq.~(\ref{eq:sPOD}) can be adopted to approximate the instantaneous streamfunction field, which is required to solve Eq.~(\ref{eq:Bouss2}) for the temperature in FOM space. On the other hand, the solution of the FOM (i.e., Eq.~(\ref{eq:Bouss2})) along with Eq.~(\ref{eq:beta}) feeds the ROM solver with $\beta$ values. This constitutes a \emph{two-way} ROM-FOM coupling problem, in contrast to the \emph{one-way} coupling in the aforementioned 1D Burgers example. We also highlight that the computational cost of the projection step (i.e., Eq.~(\ref{eq:beta})) is minimal compared to the solution of Eq.~(\ref{eq:Bouss2}).}

\textcolor{rev}{For demonstration, we consider a strong-shear flow exhibiting the Kelvin-Helmholtz instability, known as Marsigli flow or lock-exchange problem. The physical process in this flow problem explains how differences in temperature/density can cause  currents to form in the ocean, seas and natural straits. For example, Marsigli discovered that the Bosporus currents are a consequence of the different water densities in the Black and Mediterranean seas\cite{soffientino2005bosporus}. Basically, when fluids of two different densities meet, the higher density fluid slides below the lower density one. This is one of the primary mechanisms by which ocean currents are formed \cite{gill1982atmosphere}.}

\textcolor{rev}{The problem is defined by two fluids of different temperatures, in a rectangular domain $(x,y) \in [0,8] \times [0,1]$ with a a vertical barrier dividing the domain at $x=4$, keeping the temperature, $\theta$, of the left half at $1.5$ and temperature of the right half at $1$. Initially, the flow is at rest (i.e., $\omega(x,y,0) = \psi(x,y,0) = 0$), with uniform temperatures at the right and left regions (i.e., $\theta(x,y,0) = 1.5 \ \forall \ x\in [0,4]$ and $\theta(x,y,0) = 1 \ \forall \ x\in (4,8]$). Free slip boundary conditions are assumed for flow field, and adiabatic boundary conditions are prescribed for temperature field. Reynolds number of $\text{Re} = 10^4$, Richardson number of $\text{Ri} = 4$, and Prandtl number of $\text{Pr}=1$ are set in Equations~(\ref{eq:Bouss1}) and (\ref{eq:Bouss2}). A Cartesian grid of $4096\times512$, and a timestep of $\Delta t=5\times10^{-4}$ are used for the FOM simulations. Standard second-order central finite difference schemes are adopted for the discretization of linear terms while the second order Arakawa scheme \cite{arakawa1997computational} is used to compute the Jacobian term. The evolution of the temperature field is shown in Figure~\ref{fig:BSFOM} at $t=0,2,4,8$. At time zero, the barrier is removed instantaneously triggering the lock-exchange problem, where the higher density fluid (on the right) slides below the lower density fluid (on the left) causing an undercurrent flow moving from right to left. Conversely, an upper current flow moves from left to right, causing a strong shear layer between the counter-current flows. As a result, vortex sheets are produced, exhibiting the Kelvin-Helmholtz instability.
\begin{figure}[!ht]
	\centering
	\includegraphics[width=0.9\textwidth]{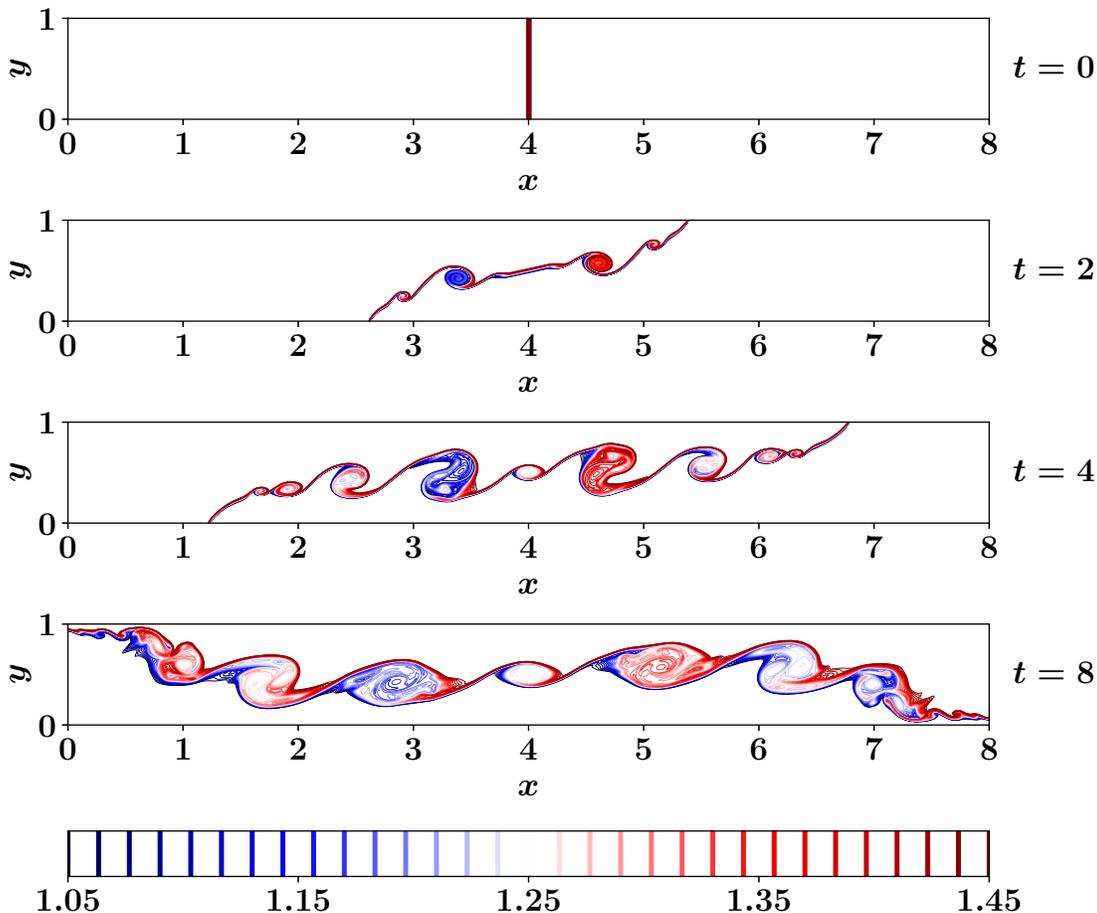}
	\caption{\textcolor{rev}{Temperature field at different time instances for 2D Boussinesq problem using $4096\times512$ grid and $\Delta t= 0.0005$.}}
	\label{fig:BSFOM}
\end{figure}}

\textcolor{rev}{Considering the dimensionality of this problem, we emphasize that the PCI approach becomes highly unfeasible. For example, with the current mesh resolution (i.e., $4096 \times 512$), the dimension of the state space of the prolonged interface is $\approx 2 \times 10^6$. Building and training of LSTMs with an input and output vector sizes of two millions become prohibitive. Even the training of convolutional neural network with such high resolutions (which are typical in fluid flow simulations) requires excessive computational resources. Therefore, we present results for ROM-FOM coupling using approaches that operate in ROM space (i.e., DPI, CPI and UPI). We note that $800$ time snapshots are stored for POD basis construction. For the Galerkin ROM solver, $r=8$ modes are retained. We also utilize the dataset of the stored $800$ snapshots for LSTM training and validation ($80\%$ randomly selected for training and the rest for validation, similar to the previous example). A two-layer LSTM with 20 cell in each layer constitutes our LSTM architecture. During the testing phase, the trained neural networks are deployed at each and every timestep. This corresponds to the application of the presented approached $16000$ times.} 

\begin{figure}[!ht]
    \centering
    \includegraphics[width=0.9\textwidth]{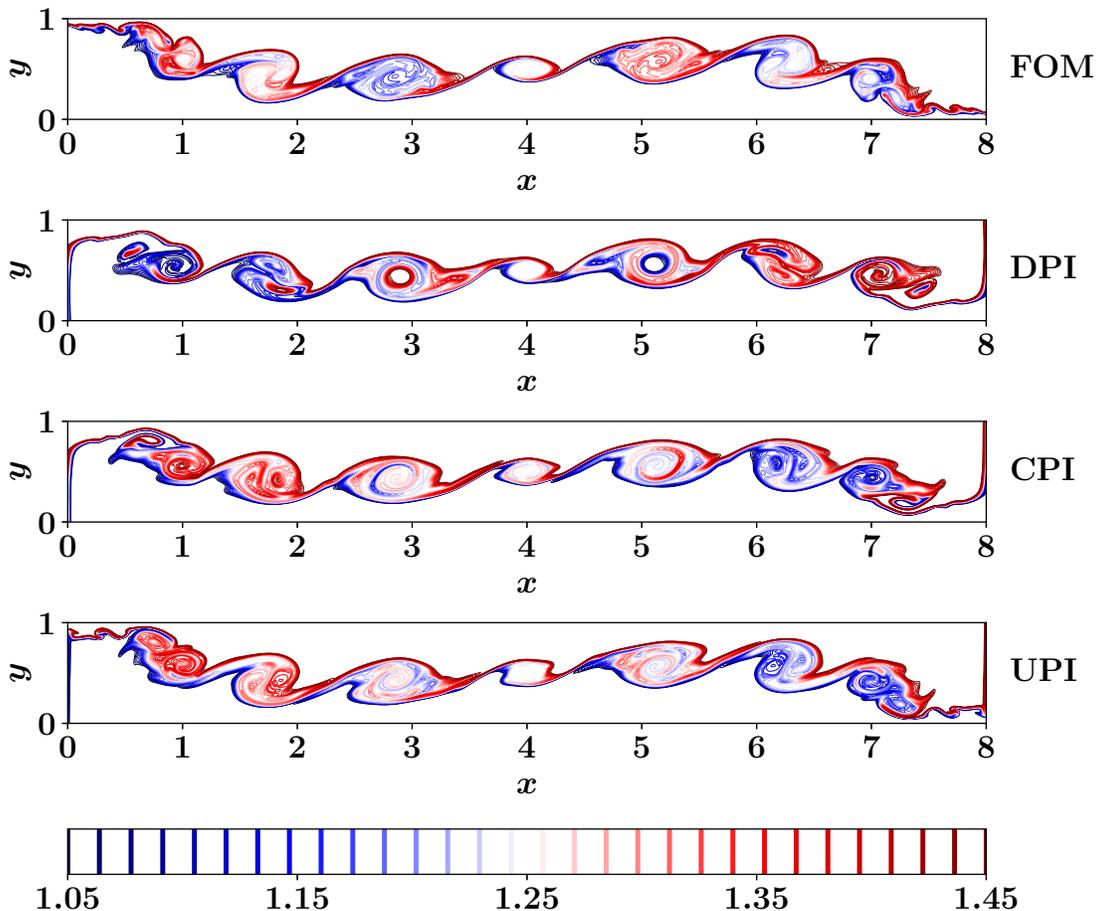}
    \caption{\textcolor{rev}{Final temperature fields as obtained from different ROM-FOM coupling approached, compared to the FOM solution. We note that the PCI becomes infeasible for higher dimensional systems.}}
    \label{fig:BScoup}
\end{figure}

\textcolor{rev}{Figure~\ref{fig:BScoup} shows the predictions of the temperature field at final time (i.e., $t=8$) computed from DPI, CPI, and UPI approaches compared to the FOM field. We emphasize that the ROM-FOM coupling results correspond to the solution of the vorticity equation with a ROM solver, which feeds the FOM solver with streamfunction to solve the temperature equation only as opposed to the FOM results which comes from the solution of both the 2D Boussinesq equations using a FOM simulation. Although the CPI results are better than those of DPI, we can observe that the fine details of the flow field are not accurately captured. That is $8$ modes are not sufficient to sufficiently represent the flow field. This is a common problem for convection-dominated flows which exhibit slow decay in the POD eigenvalues and the generated global basis functions suffer from modal deformation \cite{ahmed2019memory}. On the other hand, the implementation of the UPI approach with $r=8$ and $q=8$ recovers an increased amount of the fine flow structures that are not well-represented by the first $r=8$ modes. }

\begin{figure}[!ht]
    \centering
    \includegraphics[width=0.9\textwidth]{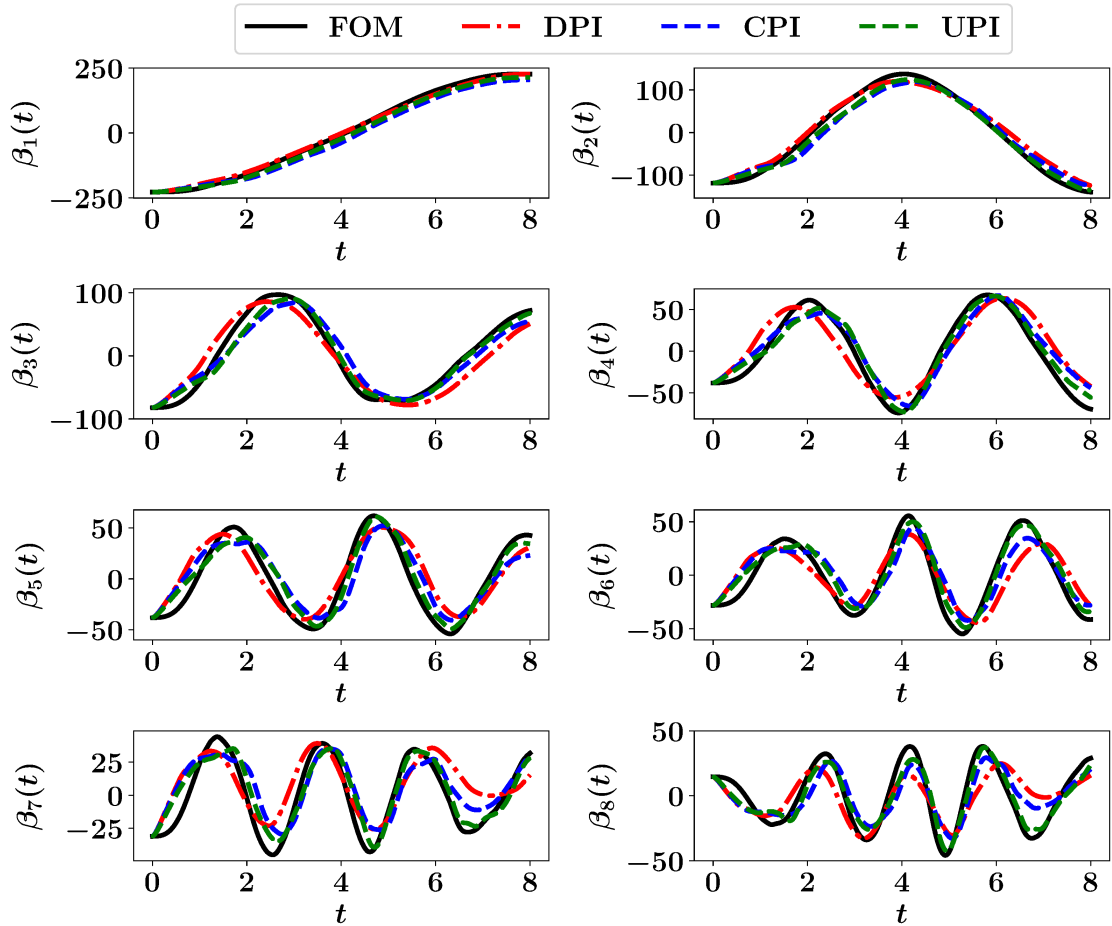}
    \caption{\textcolor{rev}{Projection of the predicted temperature fields at different times from FOM, DPI, CPI and UPI onto the POD basis function.}}
    \label{fig:BSbeta}
\end{figure}

\textcolor{rev}{Although the ROM is built for the vorticity equation only, the basis functions of the temperature fields should be generated as well to carry-out the coupling from FOM to ROM. Moreover, in order to illustrate the temporal variation of the coupling quality, we project the resolved temperature fields at different times onto their POD basis. This is depicted in Figure~\ref{fig:BSbeta}, showing the effect of different approaches on the resulting predictions of temperature fields. The FOM trajectory corresponds to the solution of both the 2D Boussinesq equations in FOM space, then projecting the obtained fields on the basis functions of $\theta$ (see Eq.~(\ref{eq:beta})). For the rest, the streamfunction fields are obtained from ROM predictions and fed into FOM solver to compute the temperature fields.}

\section{Conclusions}
We provide an interface learning approach via ROM-FOM coupling for \textcolor{rev}{multifidelity simulation environments.} This learning paradigm is built with a hybrid analysis and modeling (HAM) framework to enhance the ROM approximation of interface conditions. A demonstration with a bizonal 1D Burgers problem is considered to assess the performance of the introduced learning schemes \textcolor{rev}{for multi-component systems}. For 1D problems, we find that a prolongation followed by a machine learning correction interface (PCI) yields very good predictions. However, this might be unfeasible for 2D and 3D cases, where a correction in ROM subspace is essential. For such, a machine learning correction in ROM space followed by a prolongation interface (CPI) can produce sufficient accuracy. For complex systems where the projection error is significantly large, an uplifted prolongation interface (UPI) methodology can be adopted to recover some of the truncated scales. \textcolor{rev}{This is further illustrated using the lock-exchange problem governed by the 2D Boussinesq problem, where the ROM and FOM solvers address the vorticity and temperature equations, respectively. The coupling from ROM to FOM is represented by the 2D streamfunction fields reconstructed from the ROM solver, saving the run time for Possion equation, which is the most demanding part of an incompressible flow solver.} Owing to the relative simplicity, robustness and ease of these interface learning methods, we expect a growing number of applications in a large variety of interfacial problems in science and engineering. \textcolor{rev}{Of particular interest, this ROM-FOM coupling could be a viable method for developing next generation digital twin technologies.}   




\section*{Acknowledgements}
This material is based upon work supported by the U.S. Department of Energy, Office of Science, Office of Advanced Scientific Computing Research under Award Number DE-SC0019290. O.S. gratefully acknowledges their support. OPWIND: Operational Control for Wind Power Plants (Grant No.: 268044/E20) project funded by the Norwegian Research Council and its industrial partners (Equinor, Vestas, Vattenfall) is also acknowledged.
Disclaimer: This report was prepared as an account of work sponsored by an agency of the United States Government. Neither the United States Government nor any agency thereof, nor any of their employees, makes any warranty, express or implied, or assumes any legal liability or responsibility for the accuracy, completeness, or usefulness of any information, apparatus, product, or process disclosed, or represents that its use would not infringe privately owned rights. Reference herein to any specific commercial product, process, or service by trade name, trademark, manufacturer, or otherwise does not necessarily constitute or imply its endorsement, recommendation, or favoring by the United States Government or any agency thereof. The views and opinions of authors expressed herein do not necessarily state or reflect those of the United States Government or any agency thereof.

\section*{Data availability}
The data that supports the findings of this study are available within the article. The codes to reproduce the presented results are publicly accessible at our GitHub repository \url{https://github.com/Shady-Ahmed/ROM-FOM-Coupling}

\bibliographystyle{unsrt} 

\bibliography{manuscript}

\end{document}